# LDoS attack detection method based on traffic time-frequency characteristics[⋆]

Yu Fu[a], Xueyuan Duan[a,b,c,∗], Kun Wang[a,d], Bin Li[a]

[a]Department of Information Security, Naval University of Engineering, Wuhan, 430033, China
[b]College of Computer and Information Technology, Xinyang Normal University, Xinyang, 464000, China
[c]Henan Key Laboratory of Analysis and Applications of Education Big Data, Xinyang, 464000, China
[d]School of Mathematics and Information Engineering, Xinyang Vocational and Technical College, Xinyang, 464000, China

## Abstract

For the traditional denial-of-service attack detection methods have complex algorithms and high computational overhead, which are difficult to meet the demand of online detection; and the experimental environment is mostly a simulation platform, which is difficult to deploy in real network environment, we propose a real network environment-oriented LDoS attack detection method based on the time-frequency characteristics of traffic data. All the traffic data flowing through the Web server is obtained through the acquisition storage system, and the detection data set is constructed using pre-processing; the simple features of the flow fragments are used as input, and the deep neural network is used to learn the time-frequency domain features of normal traffic features and generate reconstructed sequences, and the LDoS attack is discriminated based on the differences between the reconstructed sequences and the input data in the time-frequency domain. The experimental results show that the proposed method can accurately detect the attack features in the flow fragments in a very short time and achieve high detection accuracy for complex and diverse LDoS attacks; since only the statistical features of the packets are used, there is no need to parse the packet data, which can be adapted to different network environments.

## Keywords

real network environment, time-frequency characteristics, low rate denial of service, attack detection

[⋆] This work was supported by National Key Research and Development Program of China under Grant 2018YFB0804104.
∗Corresponding author: Xueyuan Duan
Email addresses: fuyu0219@163.com (Yu Fu), duanxueyuan@xynu.edu.com (Xueyuan Duan), queen@xyvtc.edu.com (Kun Wang)

# 1. INTRODUCTION

With the rapid development of communication technology, the Internet continues to cover all walks of life, not only changing the mode of people's lives, but also providing a powerful boost to the development of economy and the spread of culture. However, due to the openness of network protocols, network security issues are also facing unprecedented challenges. Trojan horses, viruses and other malware are widely spread through the Internet, and network intrusions such as DoS (denial of service) attacks and zero-day attacks against application vulnerabilities never stop, which not only affect the normal operation of network services, but also cause personal privacy leakage, economic and property losses, and even threaten national security.

DoS attacks use a large amount of network attack traffic to exhaust the network bandwidth or system resources of the attacked target, making it unable to provide normal services due to overload. When the attack traffic comes from multiple attackers in different locations, it is a distributed denial of service (DDoS) attack. DDoS exploits vulnerabilities in transport, network, or application layer protocols to launch attacks [1], and due to the strategy of forging the attack source address, DDoS attacks exhibit good stealth, which is difficult to detect by traditional detection methods. difficult to be detected by traditional detection methods. Although some researchers have applied machine learning techniques to DDoS detection and achieved certain results [2][3] However, with the development of network technology, DDoS attacks have gradually changed from the original high-speed flooding attack to a low-speed sparse attack model. For example, low-speed HTTP slow read attack, the attacker initiates HTTP GET request to limit the web server's broadcast rate by setting a smaller or almost zero broadcast receive window value to achieve the purpose of occupying server resources for a long time; another example: the attacker continuously sends TCP SYN requests but does not complete three handshakes, and the server continuously sends invalid synchronization data, thus affecting the response to response to legitimate user requests. This kind of attack uses less periodic attack traffic data to occupy communication links, ports, application services, and computing resources for a long time, with the goal of consuming network bandwidth and reducing the quality of service, which we call low-rate denial of service (LDoS) attack. management mechanism, but because the attack traffic intensity is much smaller than DDoS attack traffic, LDoS attacks are more secretive, and more difficult to detect.

# 2. RELATED WORK

Kuzmanovic and Knightly[4], perhaps the first scholars to study LDoS attacks, proposed a new type of low-rate TCP directed DoS attack in 2003, which they also called since it can penetrate the network core with a small amount of traffic and destroy any target on the network thus causing great damage "sledge-hammer". Since then, many LDoS attacks have been proposed and analyzed, such as impulse attacks in the form of a series of periodic impulse signals [5], quality degradation attacks in which the effect of the attack manifests as a degradation of the quality of service [6], and thief attacks [7] that can secretly steal user data and consume computing resources on cloud platforms.

Figure 1 shows how the low-speed pulse-shaped SYN attack flow affects normal traffic. when the LDoS attack is first launched, SYN requests are sent at a high rate for a short period of time and then go into a longer hibernation period, when there is no significant change in the normal traffic in the network, however, as the number of attacks increases, normal traffic is squeezed to almost "0" and the server is unable to provide normal network services. Throughout the attack, although the peak rate of the attack flow is 50 requests per second (r/s, requests per second), the average rate is only 1.5 r/s, while the average rate of normal traffic is 4.5 r/s, and the attack traffic is masked under the normal traffic. Observing only a single instantaneous peak in the traffic, it is impossible to determine whether it is caused by the attack or by an increase in normal TCP connections. This traffic generated by LDoS attacks is almost not significantly different from normal traffic and is not intercepted by firewalls, so LDoS attacks are more covert than DoS or DDoS attacks [8], and traditional detection methods are difficult to effectively detect this attack traffic.

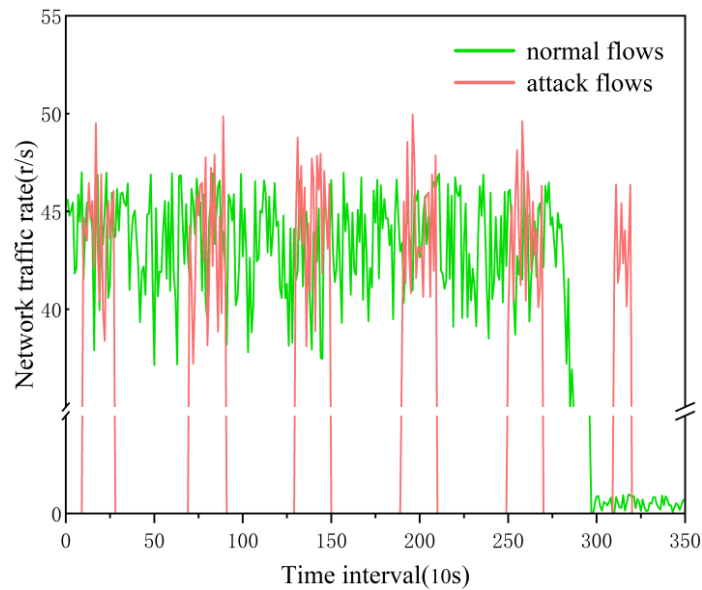

Figure 1: LDOS attack traffic graph

In this pulse-shaped LDoS attack, the attacker initiates new connections at a high rate within a short period of time and then remains inactive for a long time, and the attack traffic shows high pulse rate and periodic characteristics; while the network under attack shows the volatility of network traffic, which is mainly reflected in the characteristics of source and target IP addresses, time between new connections and traffic rate. Based on the above characteristics, LDoS attack detection methods based on traffic characteristics, machine learning and signal analysis have been proposed.

Traffic feature-based detection methods are usually used to detect abnormal fluctuation features of traffic in the attacked network, which include: queue length, connection duration, packet number, packet interval, packet size, and ACK sequence number extracted from the network[9][10].Wu et al [11] extracted three internal features of LDoS attack traffic and input to a neural network classifier for LDoS attack detection, and verified the effectiveness of their method through simulation platform and experimental network.Wu's team [12] also studied the multifractal characteristics of LDoS attack traffic [13], and proposed the multifractal trend analysis (MF-DFA) algorithm to estimate the singularity and burstiness of traffic under LDoS attack using HÖlder index, and the experimental results were consistent with the theoretical prediction. Zhang et al [14] proposed an LDoS attack detection and filtering method based on the ratio of incoming packets to the total number of packets in the flow, according to the phenomenon that normal TCP flows send fewer packets during network congestion and attack flows send more packets, which demonstrated the feasibility of the method by analyzing the data in congested routers. However, these traffic feature-based detection methods suffer from two shortcomings: first, the studies and experiments are conducted in simulated environments or data sets without validation in real networks; Secondly, the extraction of features requires very high data processing expenses and takes a long time to consume, which is only suitable for processing offline data. These two shortcomings limit the application of feature-based detection methods in real-time online detection of LDoS attacks [15].

Machine learning based detection methods usually combine machine learning methods with other algorithms. Zhang et al [16] combined principal component analysis (PCA) with support vector machine (SVM) models to detect attacks, using PCA to filter out noise interference and extract the main features of TCP flows, which were used to train SVM models. Yan et al [17] extracted the mean, variance and entropy and other features to train improved logistic regression models to detect LDoS attacks. Pérez et al [18] proposed a framework for detecting LDoS attacks in SDN environment which helps to implement various machine learning models such as decision tree, representative tree, random tree, random forest, multilayer perceptron (MLP) and support vector machine for LDoS attack detection. However, machine learning-based detection methods are equally computationally intensive in feature extraction and feature design relies on human experience, which prevents efficient detection of attack traffic with unknown features.

Many experiments have demonstrated the feasibility of using signal analysis methods to detect LDoS attacks in network traffic. Yanxiang He et al [19] designed a LDoS attack detection system using wavelet multiscale analysis to transform network traffic into fifth-order wavelet coefficients. Agrawal et al [20] used a power spectral density approach to identify low-speed LDoS attacks in cloud environments, using data collected in the time domain, using Fourier transform to the frequency domain, and calculating the power spectral density values, and if the power spectral density is concentrated in the low frequency band, it is classified as an attack. Brynielsson [21] tried to detect LDoS attacks on application servers using spectral analysis based on the continuous connection feature in the HTTP protocol. Although the spectral classification method is effective for the detection of LDoS attacks, it causes high computational overhead due to signal transformation and has a high level of false positives. In the time domain, Tang et al [22] combined Hilbert spectrum and Pearson correlation coefficient to detect LDDoS attack packets within a small-scale detection window. Zhang et al [23] proposed an adaptive detection scheme based on advanced entropy (AEB) that enables the detection of unknown attacks, and although these methods are simple and fast, they all require that the volatility of the traffic data being detected in the time domain must not be excessive, otherwise the detection effect is severely degraded, which is obviously incompatible with the real network situation.

The existing LDoS attack detection methods have two problems: first, the LDoS attack traffic used in the study is simulated and generated by using a network simulator through periodic signal superposition, which is pure and single; while the actual attack traffic is dynamically generated by the attack program, which is complex and variable. Second, the existing public datasets, although collected from real networks, the traffic data characteristics do not match the characteristics of LDoS attack traffic in the real network environment at present. For example, the CIC-2019 [24] and CIC-2017 datasets, in which the CIC-2019 dataset is mostly high-speed continuous DDoS attacks, which are easy to detect; while the CIC-2017 dataset, although it contains several types of low-rate DoS attacks, forwards more than 150 packets per second, while the rate of LDoS attacks nowadays can be as low as only a dozen packets per second or even lower. Third, most of the research experiments are conducted in experimental networks or simulation platforms, and the research results are not validated by real network environments. For example, to study a model for detecting LDoS attacks on Web servers, smooth FTP traffic is used as the background, which achieves ideal results, but if the model is deployed in a real network environment, the detection effect will be unpredictable.

In this paper, we propose a method for online detection of LDoS attacks based on time-frequency features for real network environments, in response to the problems of traditional anomalous traffic detection methods with large computational expenses and time consumption, a single type of LDoS attack stream generated by simulation, and the disconnection between the experimental background and the real network environment. Based on the real network as the background, the method realizes the end-to-end detection mode of network traffic crawling, data pre-processing, feature learning, anomaly discovery and result output.

## 3. DATA ACQUISITION AND DESIGN

### 3.1 Data Acquisition

**(1) Storage system for network traffic acquisition**

Due to the limitations of experiments using web simulators or publicly available datasets, the data used in this study were obtained from real Web service networks, preprocessed, and constructed for model training and performance evaluation. To obtain all the traffic exchanged in the Web server, a high-performance traffic acquisition and storage system is built, using Wireshark as the traffic acquisition software and a large-capacity disk array for permanent storage of the data; in addition, to alleviate the speed difference between traffic acquisition and disk storage, a high-speed RAM cache set is used to absorb the instantaneous peak data of the traffic to ensure the integrity of the collected information. Figure 2 shows the basic architecture of the network traffic collection system. The traffic acquisition storage system is deployed on the nearest switch to the web server, with the Web server traffic acquired via port mirroring.

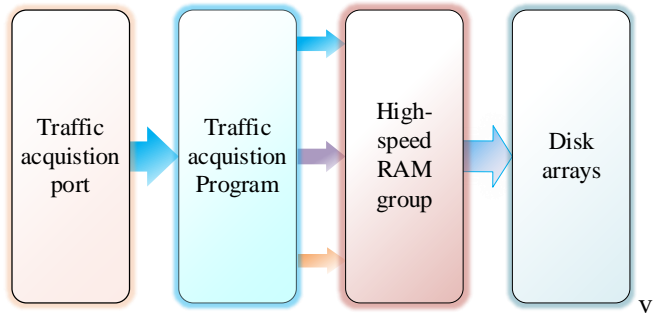

**Figure 2: Network traffic acquisition and storage system architecture diagram**

**(2) Attack traffic acquisition**

In order not to affect the normal operation of the network, the attack traffic cannot be transmitted on the Web server network, but can only be generated and acquired in a physically isolated network. Therefore, an attack traffic collection LAN is designed, and its topology is shown in Figure 3.

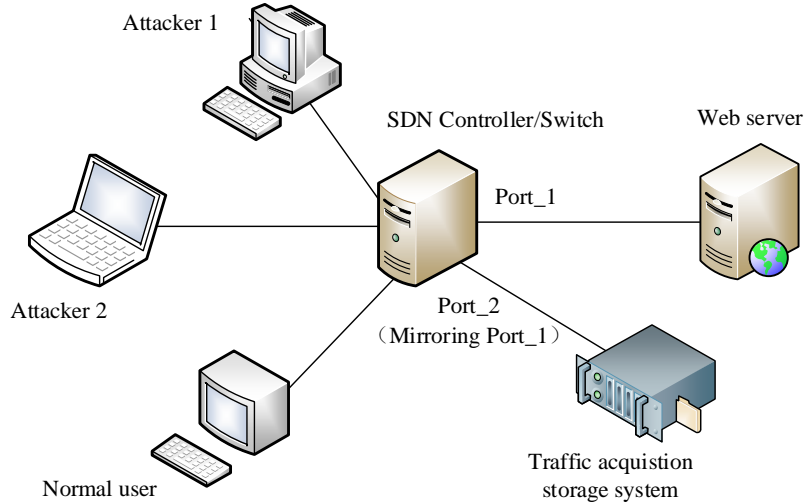

**Figure 3: Attack traffic acquisition network topology diagram**

The LAN consists of five hosts, four of which are installed with Linux and one with Windows. One Linux host runs an OpenSwitch as an OpenFlow switch, and a Pox controller as an OpenFlow controller to create an SDN environment; the southbound interface is a TCP channel with a bandwidth of 1Gbps and the OpenFlow v1.3 protocol is used for communication between the switch and the controller. 1 Linux host acts as a separate web server; 1 Linux host acts as a normal user to access the web server normally; A Linux host and a Windows host perform LDoS attacks on the web server by running different attack programs. The client and server machines are connected to the OpenFlow switch, and the specific configuration of each host is shown in Table 1.

**Table 1  LAN Host Configuration to Generate LDoS Attack Traffic**

| Host Name | CPU model(frequency) | Memory capacity and frequency | System Version (Kernel) | Running programs |
|---|---|---|---|---|
| SDN Controllers/ Switches | Intel Core i7-8750 （2.21GHz） | 32GB(2666MHz) | Ubuntu 18.04 LTS(Linux 4.15.0) | OpenSwitch/Pox controller |
| Servers | Intel Core i7-12700H （2.3GHz） | 32GB(4800MHz) | Ubuntu Server 18.04 LTS(Linux 5.2.4) | Apache 2 (Apache for Ubuntu 18) |
| Attacking host 1 | Intel Core i7-8750 （2.21GHz） | 32GB(2666MHz) | Windows 10(21H2) | Pwnloris(Httpbog) |
| Attacking host 2 | AMD Ryzen R7-4800H （2.9GHz） | 16GB(3200MHz) | Ubuntu 18.04 LTS(Linux 4.15.0) | Slowloris/ Torshammer/ Hping/ Slowhttptest |
| General users | Intel Core i7-8565U （1.80GHz） | 16GB(2133MHz) | Ubuntu 16.04 LTS(Linux 4.4.34) | Python program for accessing the server normally |

Storage system for network traffic acquisition capture all traffic flowing through the server from the mirror port of a university's web server. Four instances of the client Python program running continuously on the normal user machine communicate with two instances of the server program with different ports to generate normal HTTP traffic of different sizes and different arrival times. The attack traffic is generated by the attack programs running on each attacking host respectively, and the types of attack traffic include various low-rate TCP SYN attacks, HTTP slow read attacks, etc. In order to implement SYN low-rate attacks in small time intervals, we design each attack for 50 seconds, and then sleep for 100 seconds, and the attack occurs only in the first 0.1 seconds of each second attack mode, each attack program running net time of 60 minutes. Figure 4 gives a schematic diagram of the time cycle of the attack program running and sleep.

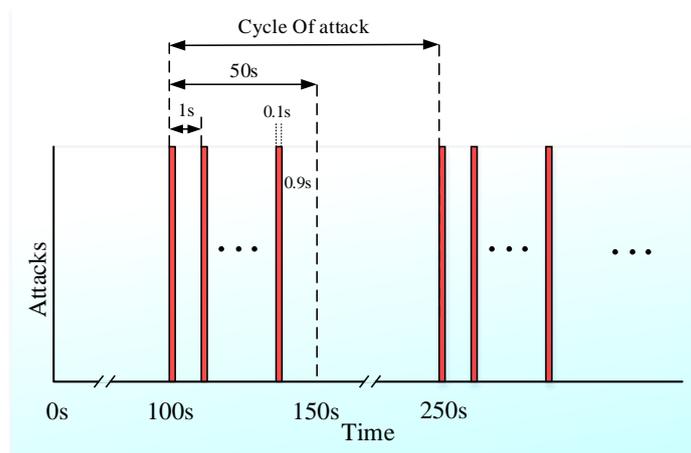

**Figure 4: Schematic diagram of the operation cycle of the attack program**

The slow-read attacks generated by the Slowhttptest program do not use the pulse-shaped attack pattern. By excluding the traffic data of the dormant period from the collected traffic, we can get six types of attack traffic with normal traffic as background, all of which are 60 minutes long. Compared to the pure attack traffic without background, the traffic we designed is more complex and more difficult to detect.

**(3) Normal traffic collection**

Figure 5 is a simplified network topology diagram containing the traffic acquisition system. The network traffic acquisition storage system is connected to the mirror port of the Web server port to capture all the traffic data of the server's external interaction. We select from the acquired traffic data for 360 consecutive minutes without network abnormalities as normal traffic, because this collection period the network can provide normal services, indicating that there is no obvious abnormality in the network, has ruled out any denial-of-service attacks, perhaps there are a small number of other types of attacks, but this attack traffic is very small, the impact on our detection of LDoS attacks can be ignored.

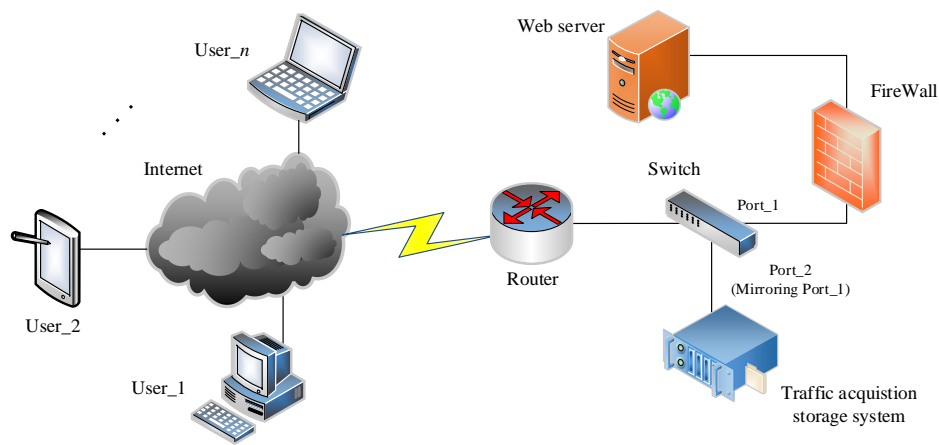

**Figure 5: Normal traffic acquisition network topology**

### 3.2 Detection data design

Since the data used in this paper are all from real networks, they need to be pre-processed, feature data selected and data set divided before they can be used as experimental data for training and detection.

**(1) Data pre-processing**

The main purpose is to remove some defective and redundant data; in addition, some traffic feature values are in the form of text, which need to be operated numerically, and the normalization of each feature value is also needed to improve the operation efficiency. Since the features selected in this study do not involve the specific content of the traffic package, and the data pre-processing is not the focus of this paper, so it is not described in detail.

**(2) Traffic feature selection**

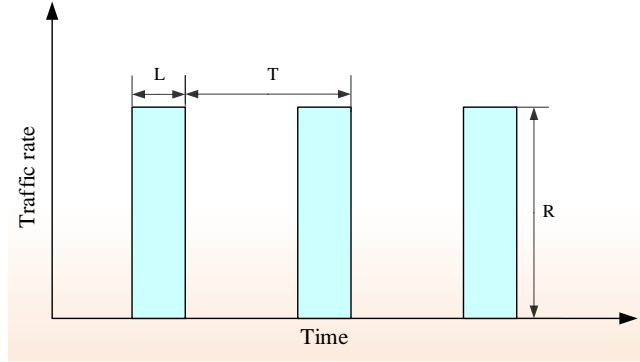

**Figure 6: General model of LDoS attack traffic**

We know that the general model of pulsed LDoS attack traffic can be represented by a triad of parameters (R, L, T), as shown in Figure 6, where R is the rate of each attack pulse, usually expressed in terms of packets transmitted per unit of time, the rate of LDoS attacks is low and does not occupy the full network bandwidth, usually only 10%-20% of normal network traffic; L is the duration of each attack pulse, and T is the attack cycle.

The traditional approach to detection is to detect LDoS attack traffic by extracting the appropriate characteristics of individual streams, which usually incurs high network overhead because of the need to track the packets of the stream. Streams are all packets with the same five elements (source IP, destination IP, source port, destination port, and traditional layer protocol). From the results of the references [25][26], it can be found that the size and arrival interval of packets within the first 2 seconds within a stream are very important for detecting DDoS. Therefore, the input sample data attempted to be constructed in this paper is 2-dimensional array consisting of two features of fixed time steps within the stream (the arrival interval time of the first number of packets in the statistical stream and the packet size). Each packet arrival is a time step, so to ensure that the set time step covers a time greater than 2 seconds, for the constructed single sample can be expressed as：

$$x_i = \begin{bmatrix} t_i^1, t_i^2, \cdots, t_i^n \\ l_i^1, l_i^2, \cdots, l_i^n \end{bmatrix}, \ t_i^j, l_i^j \in \mathbb{R}(j \in [1, n]) \quad (1)$$

where $t_i^j$ is the interval between the arrival of the *j*th packet and the *j-1*th packet in the *i*th sample, and $l_i^j$ is the size of the *j*th packet.

**(3) Data set construction and partitioning**

After pre-processing and feature selection, the stream data is cut into one stream segment every 10 seconds, so 360 minutes of normal traffic data can get 2160 segments of normal stream segments, and label them as normal samples. Each attack flow sample time is 60 minutes, so each attack flow is cut into 360 flow segments, and a total of 2160 attack flow segments can be obtained and labeled as abnormal samples. In addition, in order to facilitate the aggregation of the detection results of different modules, it is also necessary to add a unique identifier for each stream segment.

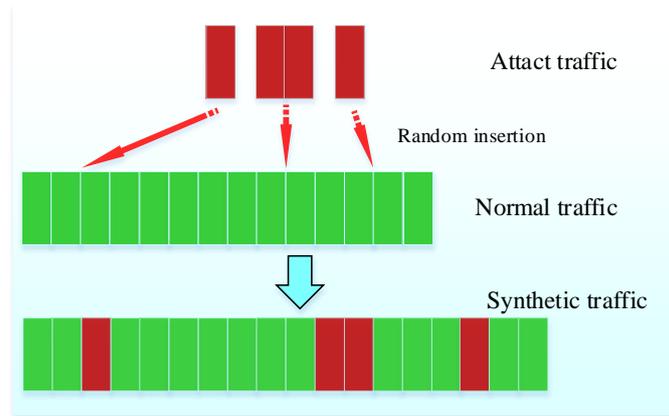

Figure 7: Synthetic traffic generation schematic

This model uses normal traffic data for training and validation. About 70% of the normal flow segments are randomly selected as the training set, and the remaining 30% of the flow segments are used as the validation set.

To construct the detection data set, one attack flow segment is randomly inserted into the normal flow each time, so that six sets of synthetic network traffic sets can be obtained by combining different attack flows with normal flows, and the detection set of each attack flow contains 2520 flow segments of 420 minutes in length, with 2160 segments for the basic normal flow segment and 360 segments for the attack flow segment.

Table 2 Information on each synthetic traffic dataset

| Name | Total traffic time (minutes) | Number of normal flow segments (segments) | Number of attack stream segments (segments) | Anomaly ratio |
| --- | --- | --- | --- | --- |
| Pwnloris | 420 | 2 160 | 360 | 0.142 9 |
| Hping | 420 | 2 160 | 360 | 0.142 9 |
| Torshammer | 420 | 2 160 | 360 | 0.142 9 |
| Slowloris | 420 | 2 160 | 360 | 10.142 9 |
| Httpbog | 420 | 2 160 | 360 | 0.142 9 |
| Slowhttptest | 420 | 2 160 | 360 | 0.142 9 |
| All-United | 480 | 2 160 | 720 | 0.25 |

Considering the diversity of attack traffic in real networks, it is necessary to use data containing multiple attack flows to test the detection capability of our proposed method for multiple attacks. Therefore, 120 attack flow segments are randomly selected from each of the six attack flows, and then randomly inserted into the normal flows to generate an "All-United" synthetic flow set containing six attack flows, which contains a total of 2880 flow segments of 480 minutes in length, including 2160 normal flow segments and 720 attack flow segments. The traffic set contains a total of 2880 flow segments of 480 minutes in length, including 2160 normal flow segments and 720 attack flow segments. Table 2 shows the relevant information of each dataset.

## 4. TFD MODEL

### 4.1 TFD Model

Numerous experiments have shown that convolutional neural networks have the efficacy of signal filters, i.e., convolutional neural network structures can be used as extractors of short-time frequency domain features of network traffic; while recurrent neural networks are better at extracting time-dependent information in sequences, i.e., time-domain features of network traffic. Therefore, in order to extract the time-domain and frequency-domain features of the traffic, a reconstruction machine based on time-domain features and a reconstruction machine based on frequency-domain features are constructed in this study, and the time-domain and frequency-domain features in the input traffic are extracted by the two reconstruction machines, and then the reconstruction sequence of the input data is generated based on the extracted features. In order to make full use of the feature information of network traffic in time domain and frequency domain, the LDoS attack traffic anomaly detection model (Time domain & Frequency domain Based Detection, TFD) based on time and frequency domain features is proposed. Figure 8 shows the overall architecture of TFD, including three functional blocks of data preparation, data reconstruction, and attack determination.

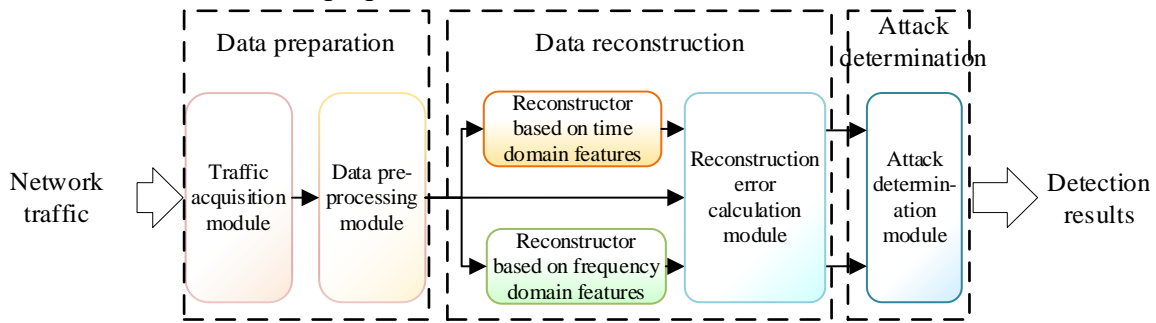

Figure 8: Flowchart of LDoS attack traffic detection based on time-frequency features

**(1) Data preparation functional group**

This group includes 2 modules of traffic acquisition and data pre-processing, which mainly capture the network traffic from the network interface and perform operations such as corresponding feature extraction and format conversion, converting it into the required form output for the next step of detection. According to the experimental findings, 16 time steps are sufficient to include the first two seconds of the flow, so the traffic data obtained from the real network, only the first 16 packets of information in the network flow studied, that is, the input data of the model is a 2-dimensional array containing 2 features of packet arrival interval and packet size for 16 time steps.

**(2) Data reconfiguration functional group**

This group consists of three modules: reconstructor based on time-domain features, reconstructor based on frequency-domain features and reconstruction error calculation. This function group is the core component of TFD, which mainly completes the reconstruction of the input feature sequence in two dimensions, time domain and frequency domain, and calculates the deviation of the generated reconstruction sequence from the input sequence, i.e., reconstruction error, as the basis for the next attack determination.

**(3) Attack determination functional group**

This group consists of the attack determination module, whose main task is to compare the reconstruction error in both time and frequency domain dimensions calculated in the previous stage with the corresponding threshold, and to determine the flow with reconstruction error greater than the threshold as an attack flow.

**4.2 Reconstructor based on time domain features**

We proposed a reconstructor based on time-domain features, which uses LSTM auto-encoder to extract and reconstruct the features of the input traffic data. In order to fully extract the data features, a channel enhancement layer is set up between the encoder and decoder, and the specific structure is shown in Figure 9.

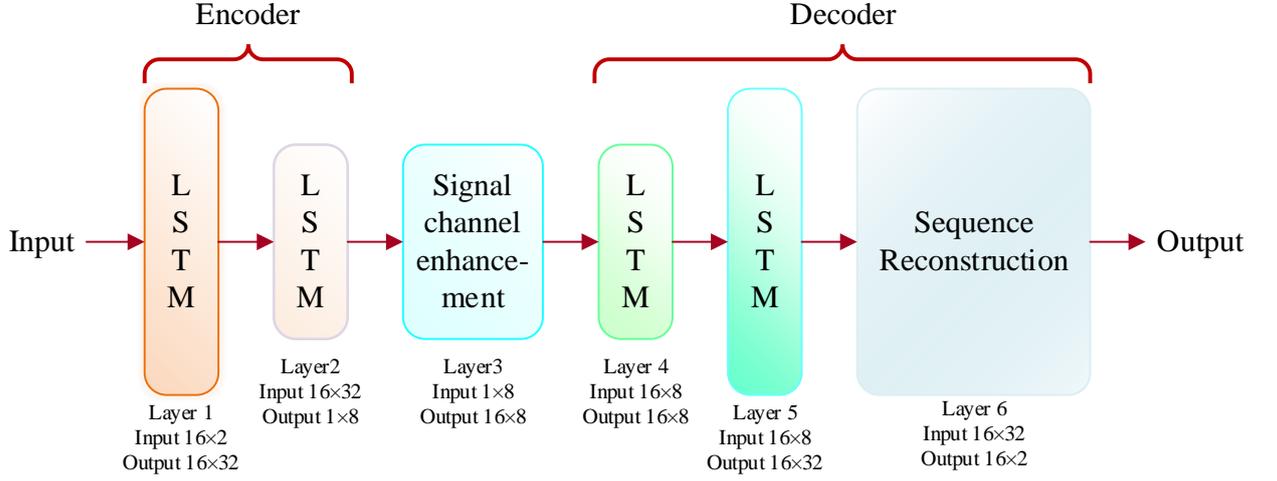

**Figure 9: Reconstructor based on time domain features**

The specific functions of each layer of the reconstructor based on time-domain features are as follows:

**(1) Encoder**

Layer 1, LSTM (32), reads the input data and outputs 32 features, each with 16 time steps.

Layer 2, LSTM (8), takes a 16×32 input from layer 1 and reduces the feature size to 8, and outputs a feature matrix of size 1×8.

Layer 3, the channel enhancement layer (16) copies the 1×8 feature matrix 16 times to form a 16×8 2-dimensional matrix as the input of the decoder layer, which can provide a richer feature representation for the decoder and is the bridge between the encoder and decoder.

**(2) Decoder**

The decoder builds layer 4 LSTM (8) and layer 5 LSTM (32) in the opposite order to the encoder, which are mirrors of layer 2 and layer 1, respectively.

Layer 6 is a fully connected layer that performs matrix multiplication between the output of layer 5 and its internal vector to generate a 16×2 output vector.

Define the encoder function in each layer as $\varphi: \mathcal{X} \to \mathcal{Z}$, which maps the input $x \in R^x = \mathcal{X}$ to $z \in R^z = \mathcal{Z}$, and the decoder layer with the function $\psi: \mathcal{Z} \to \mathcal{X}'$, which maps the input $z \in R^z = \mathcal{Z}$ to $x' \in R^x = \mathcal{X}'$.

Thus, the encoding and decoding process is expressed as：

$$z = \varphi^2(\varphi^1(x)) = \varphi^2 \circ \varphi^1(x) \qquad (2)$$

$$z' = C_{boosted}^{16}(z) \qquad (3)$$

$$x' = \psi^1(\psi^2(z')) = \psi^1 \circ \psi^2(z') \qquad (4)$$

$$f_\theta(x) = x' = \psi^1 \circ \psi^2 \circ C_{boosted}^{16} \circ \varphi^2 \circ \varphi^1(x) \qquad (5)$$

Where ∘ is the joint function, $f_\theta(x)$ denotes the function of the LSTM autoencoder defining the model, and $\theta$ is the parameters to be determined for each neuron, using the softsign activation function. The aim of our LSTM autoencoder fitting is to make the output fit the input as closely as possible, using the mean square error as the objective function:

$$Loss(\theta) = \frac{1}{m}\sum_{i=1}^{m}[x_i - f_\theta(x_i)]^2 \qquad (6)$$

where $x_i$ is the input original feature data and $f_\theta(x_i)$ is the output reconstructed data.

Using stochastic gradient descent training, the AdaGrad optimization algorithm is chosen, the learning rate in 0.000 05, the batch size in 16, and 100 complete training sessions of this reconstructor with the training set data. The training process is shown in Algorithm 1.

---

**Algorithm 1:** Training of reconstructor based on time domain features

---

Hyperparameters:
$epochs = 100$ // Number of complete training
$batch\ size = 16$ //Batch size
$\alpha = 0.000\ 05$ // Learning Rate
**Input:** Network traffic training data $X$ and label $Y$
**Output:** Model Parameter $\theta$
  for $i$ in range($epochs$)
    $g_\theta = 0$   // Resetting gradient parameter
  for $j$ in range(len($X$))   //len($X$) is the number of training data
    $x_j \in X; y_j \in Y$
    $g_\theta \leftarrow \nabla_\theta(\frac{1}{m}\sum_{i=1}^{m}[y_j - f_\theta(x_j)]^2)$   // Update gradient parameter
    $\theta \leftarrow \theta + \alpha \cdot \text{AdaGrad}(\theta, g_\theta)$   // Update model parameter
  end for
  end for
  return $\theta$

---

The reconstruction error decreases with each complete training, and eventually the reconstructor converges to a steady state in which additional training does not reduce the reconstruction error. The reconstruction error of the validation set data in the steady state is used as the threshold to distinguish normal traffic or attack traffic, namely the threshold for anomaly determination.

### 4.3 Reconstructor based on frequency domain features

We know that both Fourier transform and wavelet transform use convolutional operations to realize the conversion from time domain to frequency domain, therefore, we use convolutional neural network to extract the frequency domain features in the network traffic. Figure 10 shows the structure of our proposed frequency domain feature-based reconstructor, which is built using a convolutional residual network.

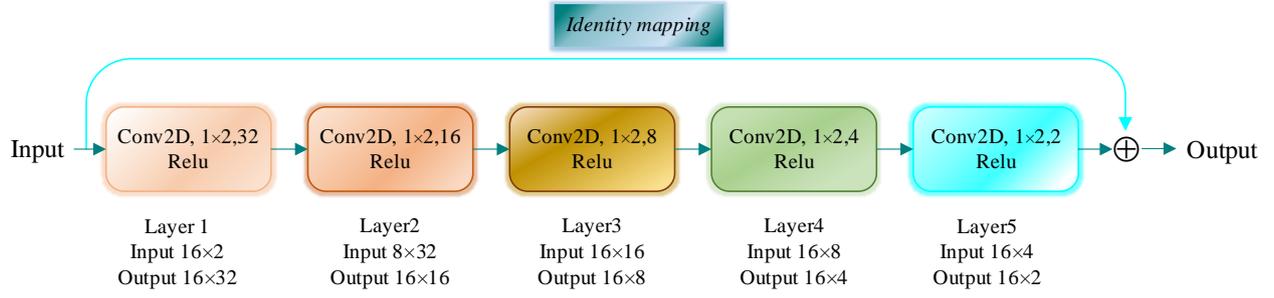

**Figure 10: Reconstructor model structure based on frequency domain features**

We design a frequency domain feature-based reconstructor consisting of five convolutional modules for extracting features of the input data in layers, denoted as Conv1 to Conv5. They all use 1 × 2 convolutional kernels with a step size of 1 and a number of channels of 1. The number of convolutional kernels in each layer is 32, 26, 8, 4, and 2, respectively, using the ReLU activation function, due to the fact that only two features in the flow are operated on, Since only two features of the traffic are operated and the feature data are small, this model does not set up pooling layer and dropout layer, and introduces residual structure to prevent the performance degradation of the neural network.

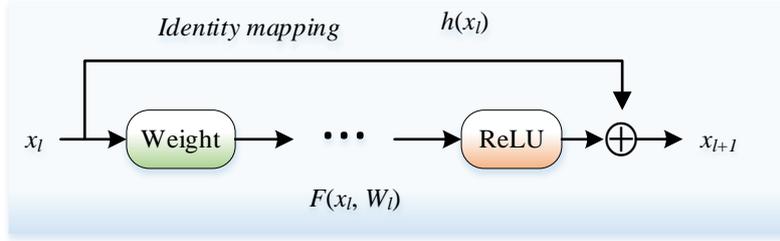

**Figure 11: Reconstructor model structure based on frequency domain features**

The residual network is composed of a series of residual blocks, and as Andreas et al [27] argue "direct mapping is the best choice", here a direct mapping of the inputs to the final design is used. Figure 11 illustrates a simple residual block structure, which can be represented as:

$$x_{l+1} = h(x_l) + \mathcal{F}(x_l, W_l) \quad (7)$$

Among them, $h(x_l)$ is the direct mapping part, which needs to be upgraded or downscaled using 1×1 convolution because the dimensions of $x_l$ and $x_{l+1}$ may be different; $\mathcal{F}(x_l, W_l)$ is the residual part, which consists of multiple convolution operations.

As the input and output dimensions of the reconstructor are the same, it can be mapped directly without dimensional change, so the final reconstructed sequence Y for the input data X after each convolution operation and residual mapping can be expressed as

$$\mathcal{F}(x, W) = \phi^5(\phi^4(\phi^3(\phi^2(\phi^1(x))))) \quad (8)$$

$$g_\vartheta(x) = x' = x + \mathcal{F}(x, W) \quad (9)$$

Where $\phi^k(\cdot)$ denotes the operation of the *k*th convolutional module, $g_\vartheta(\cdot)$ denotes the function of the convolutional residual network definition model, $\vartheta$ denotes the pending parameters of each neuron, and the cross-entropy function is used as the loss function:

$$Loss(\vartheta) = -\frac{1}{n}\sum_{i=1}^{n}[x_i \log(g_\vartheta(x_i)) - (1-x_i)\log(1-g_\vartheta(x_i))] \quad (10)$$

Where $x_i$ is the input original feature data, $g_\vartheta(x_i)$ is the output reconstruction data, trained using stochastic gradient descent method, the optimization algorithm is Adam, the learning rate is taken as 0.000 01, and the batch size is 16. 100 times of complete training is performed for this reconstructor using the training set data, and the training algorithm is analogous to Algorithm 1. The error of the model in steady state is calculated using the validation set data as the threshold for anomaly determination.

**4.4 Abnormality determination**

Using the normal data trained TFD, the reconstructor can better complete the reconstruction of the normal flow fragment, while for the attack flow fragment because it deviates from the characteristics of the normal TCP flow, the reconstructed sequence will produce a large deviation from the original input sequence, so this deviation can be compared with the reconstruction error threshold to determine whether the input is an attack flow or a normal flow. Since our purpose of detecting attack traffic is to ensure network security, we pay more attention to the recall rate (detection rate) of the attack traffic, and even the presence of appropriate false positives is tolerable, so when either of the reconstructed errors generated by the two reconstructors is greater than the corresponding reconstructed error threshold, the input is determined to be an attack traffic, that is:

$$Traffic = \begin{cases} attack & (r_T \geq R_T \text{ or } r_F \geq R_F) \\ normal & else \end{cases} \quad (11)$$

Where $R_T$、$r_T$ is the reconstruction error threshold based on the time domain features and the reconstruction error value calculated using the data to be tested, and $R_F$、$r_F$ is the reconstruction error threshold based on the frequency domain features and the reconstruction error value of the data to be tested.

# 5. EXPERIMENTS AND ANALYSIS OF RESULTS

**5.1 Experimental settings**

The hardware and software configurations of the experimental platform for model training and detection are as follows: hardware: Intel Core i9-12900F, 128GBRAM(DDR5), NVIDIA RTX3090; software: Ubuntu 18.04LTS, CUDA11.2, Pytorch1.8.

**5.2 Evaluation Indicators**

We design a fast detection method for LDoS attacks based on the TFD model with the aim of quickly discovering LDoS attack traffic from the traffic to be detected. As for the initiation phase of the attack and the type of the attack, they are not our focus, so the detection objective is finally converted into a binary classification problem. Normal traffic is defined as negative samples and attack traffic is defined as positive samples, and five metrics, Accuracy, Precision, Recall, False positive rate (FAR), and F1 value, are used to evaluate the performance of the TFD model, and these metrics are calculated as follows:

$$Accuracy = \frac{TP+TN}{TP+TN+FP+FN} \quad (12)$$

$$Precision = \frac{TP}{TP+FP} \quad (13)$$

$$Recall = TPR = \frac{TP}{TP+FN} \quad (14)$$

$$FPR = \frac{FP}{FP+TN} \quad (15)$$

$$F_1 = \frac{2 \times Precision \times Recall}{Precision + Recall} \quad (16)$$

Where TP, TN, FP, and FN indicate the interrelationship between the true and predicted results, and the specific meanings can be referred to the confusion matrix in Table 3.

Table 3 Confusion Matrix

|  | Predictive Positive | Predictive Negative |
| --- | --- | --- |
| True Positive | TP | FN |
| True Negative | FP | TN |

### 5.3 TFD model training

The two reconstructors in the TFD model have the same input data, the training process is independent of each other, and both can output results independently, so the two reconstructors are trained separately. Figure 12 shows the loss function values when the two reconstructors are trained for 100 iterations on a training set of All-United traffic data.

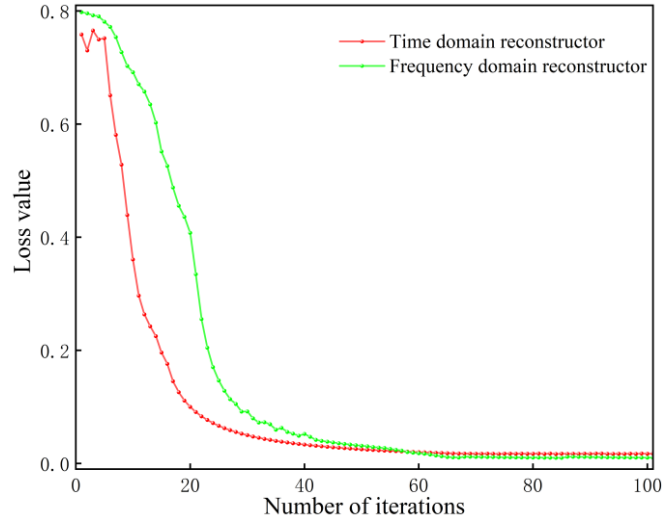

Figure 12: Anomaly recall rate for each LDoS attack traffic detection

It can be seen that the loss function values of the two reconstructors change slowly in the initial training phase, then suddenly and rapidly decrease, and stabilize after reaching a certain procedure. Relatively speaking, the time-domain-based reconstructor has a slight vibration in the loss function value at the initial training stage, but can decrease rapidly and enters a stable state first after about 50 iterations. The frequency domain-based reconstructor, on the other hand, has a stable but slow decaying loss function value at the beginning of training and stabilizes only after 70 iterations, and its steady-state loss value is smaller than that of the time-domain-based reconstructor. The reason why the time-domain-based reconstructor reaches the steady state first may be because the parameters of the model are less than those of the frequency-domain-based reconstructor model, which is easier to train and can complete the training faster, but oscillations may occur in the early training period.

In addition, because the number of samples in the dataset we use is relatively small, and the features of each sample are only a 16×2 2-dimensional matrix with few trainable parameters, the average time to complete one iteration of training on the training set for each attack traffic is no more than 7 seconds, which is converted to millisecond detection time for a single sample. In addition, since the detection target is a simple binary classification problem, the computation time to perform anomaly determination can be neglected, which means that our detection model can complete millisecond detection from data input to result output, and can be considered to be able to perform time-to-real detection of network traffic even with the addition of data collection and pre-processing time.

### 5.4 Classifier threshold setting

The error in the steady state of the two reconstructors by using the validation set as the threshold for anomaly determination during testing, combined with their performance during training, was defined as the steady state after 70 iterations, and the average of the errors generated from 71 to 100 training sessions was calculated as the threshold. In addition, to eliminate the chance in the operation, the average of the thresholds calculated for five times was taken as the final threshold of this reconstructor, as shown in Table 4.

Table 4 Detection domain values for each data set

| Validation Sets | Time Domain Reconstructor | Frequency Domain Reconstructor | |
|---|---|---|---|
| Pwnloris | 0.017 6 | 0.011 2 | |
| Hping | 0.018 6 | 0.011 6 | |
| Torshammer | 0.016 3 | 0.018 9 | Take the average of |
| Slowloris | 0.016 8 | 0.016 6 | 5 times verification |
| Httpbog | 0.073 5 | 0.018 8 | results |
| Slowhttptest | 0.082 3 | 0.063 8 | |
| All-United | 0.017 4 | 0.010 9 | |

### 5.5 Testing results

The testing experiment is divided into two phases, the first phase is to test each attack traffic individually to obtain the detection capability of the TFD model for each LDoS attack; the second phase is to test the All-United dataset consisting of multiple attack traffic to evaluate the detection capability of the model for complex attacks. A 5-fold cross-validation approach is used to calculate the mean value of the model's detection results for each dataset as the model's performance metric. The detection results for LDoS attacks alone are shown in Table 5.

Table 5  Detection performance of the model for six LDoS attacks

| Validation Sets | Accuracy | Recall | Precision | FAR | F1 |
|---|---|---|---|---|---|
| Pwnloris | 0.982 3 | 0.980 4 | 0.984 1 | 0.015 8 | 0.982 3 |
| Hping | 0.978 3 | 0.980 5 | 0.976 2 | 0.023 8 | 0.978 3 |
| Torshammer | 0.963 7 | 0.958 6 | 0.968 4 | 0.031 2 | 0.963 5 |
| Slowloris | 0.962 3 | 0.960 4 | 0.964 0 | 0.035 8 | 0.962 2 |
| Httpbog | 0.963 7 | 0.978 6 | 0.950 2 | 0.051 2 | 0.964 2 |
| Slowhttptest | 0.948 4 | 0.971 6 | 0.928 5 | 0.074 8 | 0.949 6 |
| **Average** | 0.966 5 | 0.971 7 | 0.962 0 | 0.038 8 | 0.966 7 |

As shown in Table 5, all the results are the average of five detection results, excluding the possible chance factors in computing. It can be seen that the proposed method achieves a recall rate of more than 95% for the detection of each attack traffic. Although, the false alarm rate for the Slowhttptest attack exceeds 7%, several other attacks are kept at a low level with an overall false alarm rate of 3.88%, which is acceptable in the design of detection with security alerts as the goal. The best detection indicator for Pwnloris attacks, Pwnloris is actually an upgraded version of Slowloris with more obvious features.

Secondly, the detection of Hping attacks was also good, and the reason for this was analyzed, as the Hping program was originally used for flooding attacks of DDoS. In this experiment, for the effect of low-speed attacks, Hping is set to generate attacks only in the first 0.1s interval per second, and the attack packet size is fixed, which makes the Hping attack traffic has a significant periodic change characteristic, and therefore is more easily detected.

The reason for the relatively poor detection of Slowhttptest attacks should be due to the Slowhttptest generation is a slow-read attack traffic, the time span itself is relatively large, and we choose data for the first 16 time steps in each 10-second stream segment features, resulting in the selection of some data just to deal with Slowhttptest attack traffic "silent period" so that no obvious data features, which adversely affects the detection.

The TFD model performs well in detecting individual attack traffic, but in order to study the ability of the proposed method to cope with complex attacks, it is also necessary to test the All-United data set containing multiple attack traffic, and after training the model using the data in the training set, the threshold value is derived using the validation set data, and the same 5-fold cross-validation method is used for the test set data, with the arithmetic of each result The average value is the final result, as shown in Figure 13, where the average accuracy is 0.935 8, the average precision 0.936 3, the average recall 0.940 7, the average false alarm 0.059 2, and the average F1 value 0.938 4.

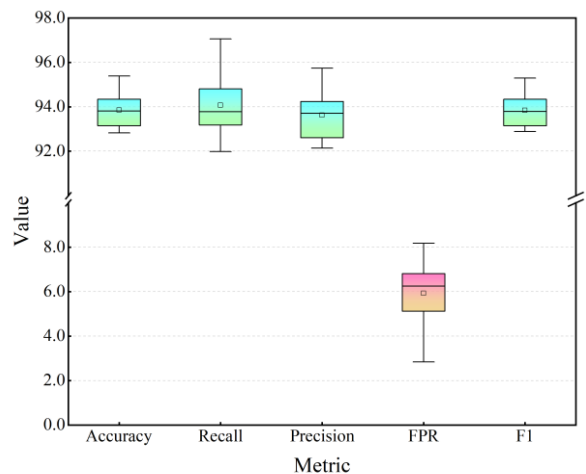

Figure 13: TFD detection metrics on the All-United attack traffic dataset

Figure 14 shows the recall rate of the TFD model for each LDoS attack type, and it can be seen that the recall rate of detecting one LDoS attack alone can reach more than 95%, and the detection rate of multiple attacks also reaches 94%. In addition, the recall here is calculated in terms of flow segments, and an attack will generate multiple flow segments, so the probability of this attack being detected will be close to 100%, and basically there will be no problem of undetected attacks, so our proposed method is effective on the dataset we designed.

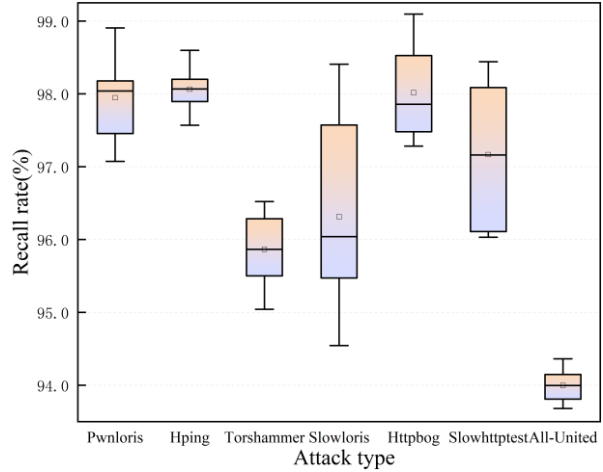

**Figure 14: Anomaly recall rate for each LDoS attack traffic detection**

To verify the adaptability of our proposed method to heterogeneous network traffic data, it is tested respectively on five publicly available datasets, including NSL-KDD, DARPA2000, ISCX2016, CICDDoS2019, and UTSA2021, where:

NSL-KDD dataset is the most commonly used dataset in the field of network traffic anomaly detection research, including ping-of-death, syn flood, smurf and other resource-consuming attacks, NSL-KDD is a dataset generated based on the improvement of KDD-CUP-99, which removes the redundant data in the KDD-CUP-99 dataset and makes an appropriate selection of the ratio of normal and abnormal data, with a more reasonable distribution of the number of test and training data.

The DARPA2000 dataset [28] is a standard dataset in the field of network intrusion detection and is one of three separate datasets in the DARPA dataset. Unlike DARPA 1998 and DARPA 1999, the DARPA2000 dataset focuses on attack traffic for Windows NT and adds internal attack and internal eavesdropping data.

The ISCX2016 low-speed denial-of-service attack dataset [29] is a dataset generated in a simulation environment, where the developers obtained eight different application-layer DoS attack traffic by building web servers such as Apache Linux, PHP5, and Drupal v7, and mixed them with the normal traffic from the original ISCX-IDS dataset to form an LDoS attack traffic dataset.

The CICDDoS2019 dataset [25] is a real-world DDoS attack-like dataset. The latest DDoS attack procedures, including reflective attacks, are used to simulate the generation of attack traffic. There are 50063112 samples in the dataset, among which there are 50006249 DDoS attack samples and 56863 normal samples, with very few normal samples, so the dataset is used with the choice of loading normal traffic from external sources or selecting only some attack samples.

The UTSA2021 dataset [30] is a set of normal and attack traffic at different rates generated using the DNS network testbed, mainly including multiple rates of TCP SYN flood attacks, HTTP slow read and slow acquisition attacks, and in this study a subset of Syn50 with an attack peak of 50r/s is used to participate in the validation.

Since we design the detection method for LDoS attacks at the transport and application layers, each data set needs to be processed to extract DoS traffic and normal traffic to form a new data set before conducting detection experiments, and then feature extraction is performed on the new data set to form the required data structure for our detection. For larger datasets, such as CICDDoS2019, only about 1% of partial attack samples are selected to improve computing efficiency.

Since the network configuration environment of each dataset has a large variability, the model TFD is trained using some normal samples from each dataset and the threshold value for anomaly determination is calculated before conducting the detection, and then the detection set containing both normal and anomalous samples is tested with the recall rate, accuracy rate and F1 value as the detection index, and the specific results are shown in Figure 15, which shows that we Concerned about the attack recall rate, basically maintain above 90%, which reached more than 98% on NSL-KDD, DARPA2000, and 96% on CICDDoS2019, UTSA2021, while the recall rate of 91.7% on ISCX2016 is relatively low. To analyze the reason for this is that the attack traffic in the three datasets NSL-KDD, DARPA2000, and CICDDoS2019 are DoS or DDoS attacks, which have more obvious statistical characteristics than LDoS in terms of packet interval and packet size, and thus can be easily identified. In the UTSA2021 dataset, we chose the attack peak of 50r/s subset Syn50, which is more challenging than detecting attack samples from a dataset with higher attack rates, and then the 96% recall is still a good result. The reason for the relatively low recall of ISCX2016 is probably due to the fact that the attack samples are generated in a different network environment than the normal samples, such that the TFD model trained using the normal samples was more rough and difficult to detect small changes in the statistical features.

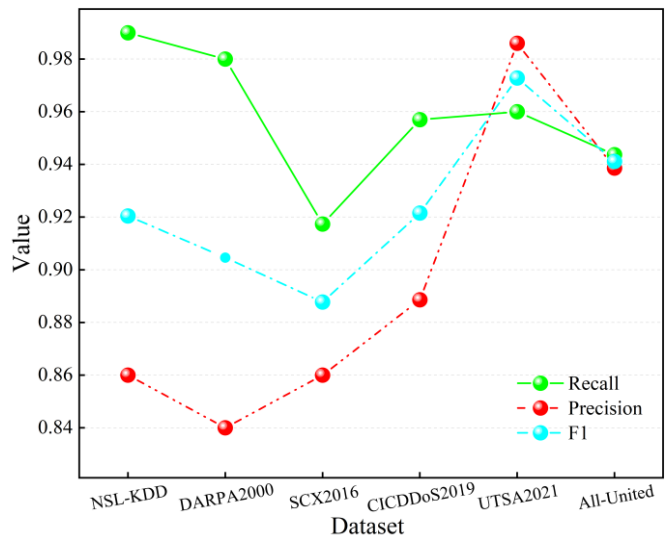

**Figure 15: TFD model for performance detection of DOS attacks in each dataset**

The accuracy of the model on the NSL-KDD, DARPA2000, ISCX2016, and CICDDoS2019 datasets did not achieve as excellent results as the recall, and the analysis may be due to the fact that we chose relatively few feature vectors and simple model results, which required much less training parameters compared to large deep network models, using sufficient samples for training. The overfitting problem occurs. This makes the model more "demanding" in determining normal traffic, so that some normal samples are mistaken for attack samples.

The UTSA2021 dataset performs the most spectacularly, even surpassing the All-United dataset we designed. The reason for this is probably due to the network traffic collection environment. The normal and attack samples of the UTSA2021 dataset are generated and collected in the same network environment, so they have good isomorphism and can be trained to produce more "pure" classifiers. In contrast, the normal samples of the All-United dataset are collected from the real network environment, which are inevitably disturbed by external conditions and produce "impurities" during the collection process. The attack samples are collected from experimental platforms with similar topology and are less subject to external interference. The model trained with the "impurity" data is a "rough" model, which may ignore the small differences between the feature data and cause misjudgment of the sample type.

The TFD model shows strong adaptability on several datasets, achieving a recall of 91.7% even on ISCX2016. Since we conduct detection experiments with stream fragments, an attack stream is composed of multiple stream fragments, which makes the detection probability of the attack stream will be much higher than the detection probability of the stream fragments, therefore, the TFD detection model we designed can well meet the original design intention of targeting the discovery of attack samples.

## 6. CONCLUSIONS

For the traditional LDoS attack detection method, it needs to extract more features, consumes more resources, cannot meet the demand of real-time online detection, and the problem that the experimental environment is too different from the real network environment. A method of attack traffic detection (TFD) based on the time-frequency domain features of network traffic is proposed, and only the arrival interval and size of the first 16 packets of the network traffic segment are selected as feature data. The TFD is modeled using normal traffic feature data. The TFD learns the spatial distribution of normal traffic features and can accurately reconstruct the normal traffic, while reconstructing the attack will generate a large reconstruction error, and the model determines the attack based on this error.

The experimental results show that the proposed method can quickly process the traffic feature data obtained from the real network environment and accurately detect the contained multiple attack traffic features with a recall rate of more than 94%, which has the capability requirement of real-time online detection. In addition, although the experimental environment is based on the background of Web services in the campus network, the proposed method can be used in principle for the detection of other LDoS attacks against HTTP servers or HTTPS servers with strong generalizability because the feature data only rely on the statistical information of packet interval and packet size in the network traffic, without the need to manually design the features, let alone analyze the specific contents of the traffic.